\documentclass[12pt]{article}
\usepackage[dvips]{graphicx}
\usepackage{epsfig,cite}
\usepackage {amsmath}
\usepackage{amssymb}
\usepackage{slashed}
\usepackage{cancel}
\usepackage{lscape}
\usepackage{axodraw4j}

\def\beq{\begin{equation}}
\def\eeq{\end{equation}}
 
\begin{document}

\begin{titlepage}
\pagestyle{empty}
\rightline{UMN--TH--3311/13, FTPI--MINN--13/39}
\vskip +0.4in

\begin{center}
{\large {\bf The Moduli and Gravitino (non)-Problems in Models with Strongly Stabilized Moduli}}
\end{center}

\begin{center}
\vskip +0.25in
{\bf Jason L. Evans$^{1,2}$, Marcos A.G. Garcia$^2$ and Keith A. Olive$^{1,2}$}
\vskip 0.2in
{\small {\it
$^1${William I. Fine Theoretical Physics Institute, School of Physics and Astronomy,\\
University of Minnesota, Minneapolis, MN 55455,\,USA} \\
$^2${School of Physics and Astronomy,\\
University of Minnesota, Minneapolis, MN 55455,\,USA}
}}


\vspace{0.8cm}
{\bf Abstract}
\end{center}
{\small In gravity mediated models and in particular in models with strongly
stabilized moduli, there is a natural hierarchy between gaugino masses, the gravitino mass and moduli masses: $m_{1/2} \ll m_{3/2} \ll m_\phi$. Given this hierarchy, we show that 1) moduli problems associated with excess entropy production from moduli decay and 2) problems associated
with moduli/gravitino decays to neutralinos are non-existent. Placed in an inflationary context,
we show that the amplitude of moduli oscillations are severely limited by strong stabilization.
Moduli oscillations may then never come to dominate the energy density of the Universe.
As a consequence, moduli decay to gravitinos and their subsequent decay to neutralinos
need not overpopulate the cold dark matter density.
}


\vfill
\leftline{October 2013}
\end{titlepage}


\section{Introduction}

The presence of light weakly-interacting fields (moduli) in the early universe has problematic consequences in cosmology \cite{prob1,prob2,enq}. Their late decay implies that these moduli may eventually dominate the energy density of the universe, redshifting as nonrelativistic pressureless matter. When they decay, they generally generate too much entropy, diluting any baryon asymmetry generated at earlier times, while failing to reheat the universe sufficiently to restart nucleosynthesis. Furthermore,
their late decays may lead to the overproduction of the lightest supersymmetric particle (LSP) \cite{myy,kmy}.

A well known example of this problem appears in the context of the Polonyi model of soft supersymmetry breaking in $\mathcal{N}=1$ supergravity \cite{pol}. At the supersymmetry breaking minimum, the scalar Polonyi field $Z$ has a mass of the order of the gravitino mass, $m_{Z}\sim m_{3/2}$. Since it couples with gravitational strength to matter fields, its decay rate is $\Gamma_{Z}\sim m_{3/2}^3/M_P^2$, where $M_P\simeq 2.4\times 10^{18}$ GeV denotes the reduced Planck mass. Moreover, during an inflationary epoch, the Polonyi field will be displaced from the minimum of the potential. Generically, this displacement is of the order of the Planck scale, $\Delta Z\sim \mathcal{O}(M_P)$ \cite{glv}. The combination of a large initial displacement, a small mass and a small decay rate is at the root of the Polonyi problem. The resulting reheating temperature,
\beq
T_R(Z)\sim \frac{m_{3/2}^{3/2}}{M_P^{1/2}},
\eeq
is smaller than the temperature required by nucleosynthesis $T_{N}\sim 1$ MeV
unless $m_{3/2}\gtrsim 10$ TeV.
Even more problematic is the entropy release from the decay of Polonyi oscillations \cite{prob1}
\beq
\frac{s_f}{s_i} \sim \frac{M_P}{m_{3/2}} ,
\eeq
where $s_{i,f}$ are the entropy densities before and after decay.
This late injection of entropy would severely dilute any pre-existing baryon asymmetry.

When the Polonyi field decays into lighter supersymmetric particles, the eventual overproduction of the LSP is likely. In particular, gravitinos may be copiously produced by the decay of the Polonyi field leading to a gravitino problem \cite{modgrav}. If the gravitino is unstable, the decay products will in turn eventually decay into the LSP, generically resulting in a dark matter relic density much larger than that observed \cite{myy,kmy}. Furthermore, the gravitino decay rate is also of the order $\Gamma_{3/2} \sim m_{3/2}^3/M_P^2$. So it too can decay late and cause problems for nucleosynthesis.

Some of the problems of the Polonyi field can be solved by giving it a larger mass\cite{enq,myy,jy}. For a modulus mass $m_Z$ (or $m_{3/2}$) larger than $\mathcal{O}(10\,{\rm TeV})$, the reheating temperature is high enough to restart nucleosynthesis after decay. The late time entropy release would nevertheless dilute the results of any previous nucleosynthesis or baryogenesis. If the baryon asymmetry is generated by a very effective mechanism, such as the Affleck-Dine (AD) mechanism \cite{AD}, which can generate a baryon-to-entropy ratio as large as $\mathcal{O}(1)$, the resulting increase in entropy could provide the necessary dilution factor to yield the observed baryon asymmetry \cite{gmo}. Other potential solutions
to the Polonyi problem have also been discussed \cite{other,Lindemech}.

In this paper, we will consider a strongly stabilized hidden sector with a Polonyi type superpotential as the source of soft supersymmetry breaking. This non-minimal Polonyi model was first introduced in \cite{dine}, and later used as part of the so-called O'KKLT mechanism \cite{okklt}. There are also several recent phenomenological
studies of strongly stabilized moduli \cite{Dudas:2006gr,klor,dlmmo,eioy,eno8}.  In\cite{ADinf}, the evolution of the strongly stabilized Polonyi sector was studied  in the context of a realization of chaotic inflation and the AD mechanism. In all of these models the cosmological problems are addressed by generating a hierarchy
between the Polonyi and gravitino masses, $m_{Z}\gg m_{3/2}$, effectively stabilizing the Polonyi field during inflation. Furthermore, the gravitino mass may be made hierarchically larger than the weak scale as in models \cite{Dudas:2006gr,klor,dlmmo,eioy}. Although this hierarchy between the gravitino mass and the weak scale is useful in curing some cosmological problems, it is also motivated by the large Higgs mass seen at the LHC\cite{lhch}. This hierarchy also implies a decay rate for the gravitino which is much larger than in the standard Polonyi model. In addition, strong stabilization also fixes the
vacuum expectation value (vev) of the Polonyi field to a value much smaller than the Planck scale,
thus reducing the amplitude of oscillations and hence the energy stored in the Polonyi field.
For a sufficiently large mass {\em and} restricted vev, dilution of the products of nucleosynthesis and baryogenesis can be avoided.   In addition, as we will show, the relic density of supersymmetric cold dark matter resulting from the decay of the modulus can be made consistent with current observations \cite{planck}. Unlike the scenarios in which the energy density is dominated by the modulus \cite{dm1}, the production of lightest supersymmetric particles does not proceed directly, but through the intermediate decay to gravitinos.

Our paper is organized as follows: In the next section, we describe the mechanism for strong stabilization and discuss the dominant possible decay modes of the inflaton.
In section 3, we incorporate an inflationary background to describe the evolution of the scalar fields.
Here we will derive the conditions on strong stabilization such that the Polonyi field or moduli
never come to dominate the energy density of the universe and hence lead to an insignificant increase in the entropy density. In section 4, we determine the resulting non-thermal relic density of LSPs
and show the conditions under which it is compatible with observations. Our summary and conclusions are given in section 5.

\section{Strong stabilization and decay modes}

In what follows, we will distinguish between three sectors of the theory:
the supersymmetry breaking sector characterized by a Polonyi-like field $Z$;
an inflation sector characterized by an inflaton, $\eta$; and a matter sector
characterized generically by a set of fields, $\phi$.
The strongly stabilized Polonyi sector is described by a superpotential
\beq\label{W}
W= \mu^2(Z+\nu),
\eeq
where the parameter $\nu$ is adjusted so that the cosmological constant vanishes at the supersymmetry breaking minimum. Unless explicitly noted, we will work in units where the reduced Planck mass, $M_P$, has been set to be unity. The K\"ahler potential includes a strongly stabilizing term added to the minimal term \cite{dine},
\beq\label{K}
K = Z\bar{Z}-\frac{(Z\bar{Z})^2}{\Lambda^2},
\eeq
where it is assumed that the mass scale $\Lambda\ll 1$.  For simplicity, we will denote by $Z$ both the chiral superfield and its scalar component. The scalar potential derived from the K\"ahler potential (\ref{K}) and the superpotential (\ref{W}) is given by  \cite{Fetal}
\beq\label{potential}
\begin{aligned}
V &= e^{K}(K^{Z\bar{Z}}D_{Z}W\bar{D}_{\bar{Z}}\bar{W}-3|W|^2)\\
&= \mu^4 e^{Z\bar{Z}-(Z\bar{Z})^2/\Lambda^2}\left[\frac{|1+\bar{Z}(1-2Z\bar{Z}/\Lambda^2)(Z+\nu)|^2}{1-4(Z\bar{Z})/\Lambda^2} - 3|Z+\nu|^2\right],
\end{aligned}
\eeq
where
\begin{equation}
D_{Z}W=(\partial_{Z}K)W+\partial_{Z}W.
\end{equation}
In order to obtain phenomenologically acceptable soft scalar masses, we assume that this Polonyi sector is hidden from the visible sector. Therefore, if $\phi$ denotes collectively the matter superfields, the K\"ahler potential and the superpotential are assumed to be separable
\begin{align}
K &= K(Z,\bar{Z}) + K(\phi,\bar{\phi}), \label{K_phi}\\
W &= W(Z) + W(\phi). \label{W_phi}
\end{align}
$K(Z,\bar{Z})$ is assumed to be given by (\ref{K}) and for our purposes here,
it is sufficient to assume that $K(\phi,\bar{\phi}) = \phi \bar{\phi}$.
We note that phenomenological studies \cite{dlmmo,eioy} show that the visible sector often requires a Giudice-Masiero term of the form
\begin{eqnarray}
K(\phi,\bar \phi) \supset c_H H_1H_2 +h.c 
\end{eqnarray}
where $H_1,H_2$ are the Higgs fields of the MSSM and $c_H$ is some dimensionless constant.

The complex field $Z$ can be parametrized in terms of
its real and imaginary parts,
\beq\label{zeta}
Z=\frac{1}{\sqrt{2}}(z+i\chi).
\eeq
In this parametrization, the supersymmetry breaking Minkowski minimum is found to be real and located at
\beq\label{z_min}
\langle z\rangle_{\rm Min}\simeq \frac{\Lambda^2}{\sqrt{6}}\ , \ \ \langle \chi\rangle =0\ , \ \ \nu\simeq \frac{1}{\sqrt{3}}\ .
\eeq
for $\Lambda\ll 1$. The supersymmetry breaking mass scale given by the gravitino mass is
\beq
m_{3/2}=\langle e^{K/2}W\rangle \simeq\mu^2/\sqrt{3} \, ,
\eeq
whereas the mass squared of both $z$ and $\chi$ are
\beq
m_{z,\chi}^2\simeq \frac{12\,m_{3/2}^2}{\Lambda^2}\gg m_{3/2}^2 .
\label{mz}
\eeq
Thus, for $\Lambda \ll 1$, we obtain the hierarchy mentioned earlier between the modulus and the gravitino.

The goldstino is the fermionic component of $Z$, $\psi_Z$. In the unitary gauge it is absorbed by the gravitino, becoming its longitudinal component, via the super-Higgs mechanism \cite{shiggs}. It is worth noting that we have also explored the scenario in which the non-minimal K\"ahler term is positive, $K = Z\bar{Z} + (Z\bar{Z})^2/\Lambda^2$. In this case, for $\Lambda\ll 1$, along the real axis ($\chi=0$), the parameter $\nu$ can be tuned to yield a Minkowski supersymmetry breaking minimum, together with an Anti-de Sitter minimum, both separated by a barrier of finite size about the origin. However, as a function of real and imaginary parts $z,\chi$, the AdS extremum is found to be the global minimum, while the Minkowski extremum is actually a  saddle point, and is connected to the global minimum in the complex direction.

The decay modes of the Polonyi field are determined by its couplings to matter and gauge fields, and to the gravitino. The interaction with matter scalars follows from the Lagrangian
\beq
\mathcal{L}_{S} = G_{i\bar{j}} D_{\mu}\phi_i D^{\mu}\bar{\phi}_{\bar{j}} - e^{G}(G_i G^{i\bar{j}}G_{\bar{j}}-3),
\eeq
with $G=K+\log|W|^2$ the K\"ahler function. Eq. (\ref{K_phi}) implies that the relevant interaction
for decay corresponds to the potential term. As noted earlier, for simplicity, we consider a minimal K\"ahler potential for matter fields. Under the assumption that the vevs of the matter fields are either zero or at most of order the weak scale, we set $K_i , W_i\ll 1$. The scalar potential can then be Taylor expanded to give the two body decay coupling,
\beq
\mathcal{L}_{S,2}= \sqrt{3}m_{3/2}(m_{3/2} - \bar{W}(\bar{\phi}) )\,Z\phi_i\bar{\phi}_{\bar{i}} + h.c. + \mathcal{O}(\Lambda^2).
\eeq
The resulting decay rate is suppressed by $\Lambda\ll1$. Restoring the Planck mass $M_P$, the width is
\beq
\Gamma(z\rightarrow\phi_i\bar{\phi}_{\bar{i}}) \simeq \frac{\sqrt{3}\Lambda}{32\pi}\frac{m_{3/2}^3}{M_P^3}.
\eeq
A further expansion reveals that the three body decays to matter scalars are determined by the Yukawa couplings of the matter fields,
\beq
\mathcal{L}_{S,3}=\sqrt{3}m_{3/2} W_{ijk} \bar{Z}\phi_i \phi_j \phi_k + h.c. + \mathcal{O}(\Lambda^2).
\eeq
In this case the decay rate is enhanced by $\Lambda$,
\beq
\Gamma(Z\rightarrow \bar{\phi}_i \bar{\phi}_j\bar{\phi}_k)\simeq \frac{3\sqrt{3}|W_{ijk}|^2}{256\pi^3\Lambda}\frac{m_{3/2}^3}{M_P}.
\eeq

The interaction of $Z$ with matter fermions is determined by the kinetic and mass terms of the supergravity Lagrangian,
\beq
\begin{aligned}
\mathcal{L}_{F}= &\ \frac{i}{2}G_{i\bar{j}}\bar{\chi}^i_R \gamma^{\mu}D_{\mu}\chi^j_R + \frac{i}{2}(-G_{ij\bar{k}}+\frac{1}{2}G_{i\bar{k}}G_j)\bar{\chi}_{R}^i\gamma^{\mu}D_{\mu}\phi^j\chi^{k}_R\\
&\ - \frac{1}{2}e^{G/2}(-G_{ij}-G_i G_j + G_{ij\bar{k}}G^{l\bar{k}}G_l)\bar{\chi}^i_R \chi^j_L +h.c.
\end{aligned}
\eeq
After the Goldstino component is subtracted out, the interactions for the two body decays $Z\rightarrow\bar{\chi}^i\chi^j$ are found to be given by
\beq
\mathcal{L}_{F,2}=i\frac{\sqrt{3}}{4}\bar{\chi}^i_R\gamma^{\mu}\partial_{\mu}Z\chi_{iR} +  \frac{1}{\sqrt{3}m_{3/2}}\bar{\chi}^i_R\left(ZW_i W_j \right)\chi^j_L+ \frac{\sqrt{3}}{4} m_{3/2} c_H Z{\bar{\tilde H}_L} \tilde  H_R + h.c. + \mathcal{O}(\Lambda^2)
\eeq
where ${\tilde H}^T= [\tilde  H_1^T, {\tilde H}_2^\dagger]$.  The squared amplitude for first term is suppressed by the masses of the final-state fermions, in addition to a factor of $\mathcal{O}(\Lambda)$. The amplitude of the second term is suppressed by the expectation values $\langle W_i\rangle\ll1$. The interference term between the first two terms vanishes. The third term, turns out to be the most dominant two-body decay mode to matter fields. It gives a decay width of
\begin{eqnarray}
\Gamma(Z\to \tilde H_1\tilde H_2)= \frac{3c_H^2}{256\pi}\left(\frac{m_{3/2}}{M_P}\right)^2m_Z= \frac{3 c_H^2}{256\pi}\frac{m_{3/2}^3}{M_P^2}\frac{M_P}{\Lambda}
\end{eqnarray}

Three body decays which include fermions in the final state, and which proceed through four point vertices are also suppressed by the expectation values of $W_i$.
The largest contribution to the decay into fermions, other than possibly the Higgsinos, is given by the fermion exchange diagram of Figure 1, with the rate
\beq
\Gamma(Z\rightarrow \bar{\chi}_i\chi_j\phi_k)\simeq \frac{9\sqrt{3}|W_{ijk}|^2}{2048\pi^3\Lambda^3}\log\left(\frac{m_z}{m_k}\right){m_{3/2}^3}{M_P}.
\eeq
Here we have neglected the masses of the final state fermions.

\begin{figure}
\centering
  \begin{picture}(114,114) (79,-15)\label{figure1}
    \SetWidth{1.0}
    \SetColor{Black}
    \Line[dash,dashsize=5,arrow,arrowpos=0.5,arrowlength=5,arrowwidth=2,arrowinset=0.2](80,66)(144,66)
    \Line[arrow,arrowpos=0.5,arrowlength=5,arrowwidth=2,arrowinset=0.2](144,66)(192,98)
    \Line[arrow,arrowpos=0.5,arrowlength=5,arrowwidth=2,arrowinset=0.2](144,18)(144,66)
    \Line[arrow,arrowpos=0.5,arrowlength=5,arrowwidth=2,arrowinset=0.2](192,50)(144,18)
    \Line[dash,dashsize=5,arrow,arrowpos=0.5,arrowlength=5,arrowwidth=2,arrowinset=0.2](192,-14)(144,18)
    \Text(78,72)[lb]{\Black{$Z$}}
    \Text(193,-10)[lb]{\Black{$\phi_k$}}
    \Text(193,38)[lb]{\Black{$\chi_j$}}
    \Text(193,85)[lb]{\Black{$\bar{\chi}_i$}}
    \Text(127,38)[lb]{\Black{$\chi_i$}}
  \end{picture}
\caption{Three body decay of $Z$ into a final state with two fermions}
\end{figure}
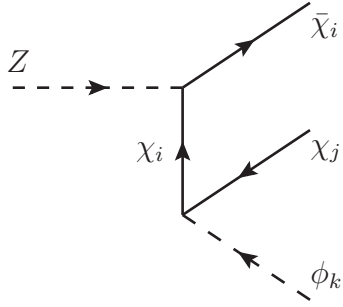

Without an explicit coupling through the gauge kinetic function, $Z$ does not decay into gauge bosons or gauginos at tree level. Nevertheless, the Polonyi field can still decay into the gauge supermultiplets through anomaly mediated effects \cite{ety,ny}. The corresponding decay rate is suppressed by a factor of $\mathcal{O}(\Lambda)$,
\beq
\Gamma(Z\rightarrow gg,\tilde{g}\tilde{g})\sim \frac{N_g\alpha^2}{256\pi^3}|K_Z|^2\frac{m_{z}^3}{M_P^2}\sim \frac{\Lambda N_g\alpha^2}{256\pi^3}\frac{m_{3/2}^3}{M_P^3}.
\eeq

In addition to the decays into matter and gauge fields, the Polonyi field can decay into gravitinos. This process is mediated by the interaction terms
\beq
\begin{aligned}
\mathcal{L}_{3/2} &= \frac{1}{8}\epsilon^{\mu\nu\rho\sigma}\bar{\psi}_{\mu}\gamma_{\nu}\psi_{\rho}G_iD_{\sigma}\phi_i + \frac{i}{2}e^{G/2}\bar{\psi}_{\mu L}\sigma^{\mu\nu}\psi_{\nu R} + {\rm h.c.}\\
&= \frac{\sqrt{3}}{8}\epsilon^{\mu\nu\rho\sigma}\bar{\psi}_{\mu}\gamma_{\nu}\psi_{\rho} \partial_{\sigma}Z - \frac{\sqrt{3}}{8}m_{3/2} Z\bar{\psi}_{\mu}[\gamma^{\mu},\gamma^{\nu}]\psi_{\nu} + {\rm h.c.} + \mathcal{O}(\Lambda^2) + \cdots
\end{aligned}
\eeq
where the ellipsis includes higher order terms in $Z$, as well as the couplings of the gravitino to scalar fields other than $Z$. The rate obtained from these couplings is enhanced by a factor of $\Lambda^{-5}$,
\beq\label{gammaz}
\Gamma(Z\rightarrow \psi_{3/2}\psi_{3/2}) \simeq \frac{3\sqrt{3}\,m_{3/2}^3 M_P^3}{\pi\Lambda^5 }.
\eeq
Alternatively, the decay rate may be computed in the goldstino picture \cite{jeong} with the same result. This rate differs from the rate computed in the standard Polonyi scenario without stabilization \cite{ety,ny}, since the interaction for the goldstino comes from the strongly stabilizing contribution to the K\"ahler potential
\begin{eqnarray}
\int d^4\theta K\supset -\int d^4\theta   \frac{|Z|^4}{\Lambda^2} \supset -2\frac{F_Z^\dagger Z^\dagger \psi_z\psi_z}{\Lambda^2} =-2\sqrt{3}\frac{m_{3/2}M_P}{\Lambda^2} Z^\dagger \psi_z \psi_z.
\end{eqnarray}
Using this and calculating the decay width we get exactly what the result above for the decays to gravitinos. The dominant channel for the spontaneous decay of the Polonyi field is therefore that to gravitinos, $\Gamma_{z}^{({\rm total})}\simeq \Gamma(Z\rightarrow \psi_{3/2}\psi_{3/2})$.

\section{Post-inflationary dynamics}

During an inflationary epoch, the scalar field $Z$ will be displaced from its true minimum given by eq. (\ref{z_min}) to smaller values. In supergravity, large masses of the order of the Hubble parameter during inflation, $H_I$, are generically induced on scalar fields, due to the exponential factor in (\ref{potential}). If $\eta$ denotes the scalar field responsible for inflation, a contribution
\beq\label{inf_corr}
\Delta V(Z) \sim e^{K(Z)}V(\eta) = cH_I^2 Z\bar{Z} + \cdots
\eeq
will typically arise, where $c\sim 3$ 
in general. If this is the case, the expectation value of $Z$ will be several orders of magnitude smaller than the true minimum, $\langle Z\rangle_{\rm inf}\ll\langle Z\rangle_{\rm Min}$. In particular, the addition of a contribution of the form of eq. (\ref{inf_corr}) to the potential (\ref{potential}) results in the vacuum expectation value during inflation \cite{ADinf}
\beq
\langle z\rangle_{\rm inf} \simeq \frac{\Lambda^2}{\sqrt{6}}\left(1+\frac{3c}{2}\frac{H_I^2\Lambda^2}{\mu^4}\right)^{-1} \simeq \sqrt{\frac{2}{3}}\frac{\mu^4}{3cH_I^2} \ll \langle z\rangle_{\rm Min}.
\eeq
At the end of inflation, the universe is dominated by the oscillations of the inflaton which leads to an
expansion rate characterized by matter domination.  During this period, the Hubble parameter decreases and the Polonyi field will adiabatically track the instantaneous minimum \cite{Lindemech} until the Hubble parameter becomes of the order of the mass of $z$; more precisely when $H=\frac{2}{3}m_z$. When this occurs, $z$ will start damped oscillations about the true supersymmetry breaking minimum (\ref{z_min}). This may occur either before or after the inflaton oscillations have decayed.
Thus the amplitude of oscillations in this strongly stabilized model is reduced relative to the
standard case by the fact that
the final vev is of order $\Lambda^2/M_P \ll M_P$.

The energy density and the Hubble parameter during the epoch where inflaton oscillations dominated universe can be written as
\begin{align}
\rho_{\eta}&=\frac{4}{3}m_{\eta}^2M_P^2 \left(\frac{R_{\eta}}{R}\right)^{3}, \label{rhoeta}\\
H &=\frac{2}{3}m_{\eta}\left(\frac{R_{\eta}}{R}\right)^{3/2}, \label{H1}
\end{align}
where $R_{\eta}$ denotes the cosmological scale factor at the onset of oscillations of $\eta$. Therefore, the oscillation of $z$ starts when the scale factor is
\beq
R_z \simeq \left(\frac{\Lambda}{2\sqrt{3}}\frac{m_{\eta}}{m_{3/2} M_P}\right)^{2/3}R_{\eta},
\label{RzReta}
\eeq
where we have used (\ref{mz}) for $m_z$.

During inflation, the imaginary part of $Z$, $\chi$, at the minimum is not displaced and so evolves to the minimum of the potential and does not oscillate thereafter. Hence, the energy density of the Polonyi field is stored in the oscillations of the real part $z$,
\beq
\rho_z\simeq \frac{1}{2}m_z^2 \langle z\rangle^2_{\rm Min} \left(\frac{R_z}{R}\right)^3.
\label{rhoz}
\eeq
The amplitude of the oscillations is therefore suppressed by $\Lambda^2$. In addition, note that for the strongly stabilized modulus we also have an enhanced mass and  therefore an enhanced decay rate. Therefore, as we show below, for a sufficiently small $\Lambda$, the cosmological problems for $Z$ are averted. The details of the evolution of $Z$ depend on whether the Polonyi field decays before or after reheating.
Some numerical results for the evolution of $z$ and $\chi$ can be found in \cite{ADinf}.

Assuming that the inflaton decays due to gravitational-strength interactions, we can parametrize the coupling by $d_{\eta}$, such that the decay rate is
\beq
\Gamma_{\eta}=d_{\eta}^2\frac{m_{\eta}^3}{M_P^2}.
\eeq
In the instantaneous approximation, the inflaton decays when $\Gamma_{\eta}=\frac{3}{2}H$, or $R_{d\eta}/R_{\eta}=(M_P/d_{\eta}m_{\eta})^{4/3}$. Comparing $R_{d\eta}$ with $R_z$,
we see that Polonyi oscillations begin before inflaton decay so long as
\beq
\left(\frac{d_\eta^2 \Lambda}{m_{3/2}}\right)^{2/3} \left(\frac{m_\eta}{M_P}\right)^2 < 1 .
\eeq
For $m_\eta \sim 10^{-5} M_P$ and $m_{3/2} \sim 10^{-15} M_P$, this condition
is valid for $d_\eta^2 \Lambda < M_P$ as we will assume. If $\Lambda$ is very
small, $Z$ will decay before the inflaton.
Very early decays of the Polonyi field will occur when $R_{dz} < R_{d\eta}$
where $R_{dz}$ is the scale factor at the time of $z$ decay.
As we will see, this condition is satisfied when $d_\eta^{2/5} \Lambda/M_P <
 10^{-6}$. We will return to this case below.

After inflaton decay, the universe is filled with the relativistic decay products, with energy density and Hubble parameter
\begin{align}
\rho_{r} & = \frac{4}{3}d_{\eta}^{-4/3}m_{\eta}^{2/3}M_P^{10/3}\left(\frac{R_{\eta}}{R}\right)^{4},
\label{rhor}\\
H_r & = \frac{2}{3}d_{\eta}^{-2/3}m_{\eta}^{1/3} M_P^{2/3} \left(\frac{R_{\eta}}{R}\right)^{2}.
\end{align}

The corresponding reheating temperature is
\beq
T_{R}=d_{\eta}\left(\frac{40}{\pi^2 g_{\eta}}\right)^{1/4}\frac{m_{\eta}^{3/2}}{M_P^{1/2}}.
\label{tr}
\eeq
where $g_{\eta}=g(T_R)$ is the effective number of degrees of freedom at reheating. For large $\Lambda$ (but still $\lesssim M_P$), the Universe may become dominated by
$z$ oscillations before they decay (as in the standard Polonyi scenario).
If this should happen, the Hubble parameter becomes
\beq
H_z  = \frac{1}{6}m_{z}\frac{\Lambda^2}{M_P^2} \left(\frac{R_{z}}{R}\right)^{3/2}.
\eeq
In this case, the scale factor at $z$ decay is
\beq
R_{dz} = \left(\frac{\pi}{6}\right)^{2/3} \frac{\Lambda^4}{m_{3/2}^{4/3} M_P^{8/3}}\, R_z
\eeq
Using (\ref{rhor}) for the energy density in radiation (subdominant), (\ref{rhoz}) for the energy density
in $z$ oscillations, and (\ref{RzReta}) to relate the scale factors $R_z$ and $R_\eta$
we can compute the entropy increase due to $z$ decays.
Taking the entropy density in radiation to be $s_{r}=4/3(g_{\eta}\pi^2/30)^{1/4}\rho_{r}^{3/4}$, and a similar
expression for the entropy density produced from $z$ decays,
we find,
\beq
\frac{s_z}{s_r} \simeq 0.05\,d_{\eta}\left(\frac{g_z}{g_{\eta}}\right)^{1/4}\left(\frac{\Lambda}{M_P}\right)^{13/2}\left(\frac{m_{\eta}}{m_{3/2}}\right)^{3/2} ,
\eeq
for the entropy ratio.
Then for our nominal values of $m_\eta \sim 10^{-5} M_P$ and $m_{3/2} \sim 10^{-15} M_P$
the entropy ratio is approximately $10^{14} \Lambda^{13/2}$.
Clearly for $\Lambda \sim M_P$, a huge amount of entropy is produced as in the original Polonyi
scenario.  However, as one can, the entropy increase is a sensitive function of the
stabilization scale $\Lambda$ and for $\Lambda \lesssim 10^{-2}M_P$,
the entropy increase becomes tolerable.

For smaller $\Lambda$, even if
 $Z$ decays after reheating, the energy density may never become dominated by $z$ oscillations. In this case,  the scale factor at the time of decay is such that $\Gamma_{z}=t^{-1}=2H_r$. With the decay width given by eq. (\ref{gammaz}), the scale factor at $Z$ decay, $R_{dz}$ is found to be
\beq
\frac{R_{dz}}{R_{\eta}}\simeq 0.9\, d_{\eta}^{-1/3}\Lambda^{5/2}m_{\eta}^{1/6}m_{3/2}^{-3/2}M_P^{-7/6}.
\eeq
This assumes that the universe is dominated by radiation  when $Z$ decays, i.e., $\rho_r/\rho_z>1$. This is valid so long as the parameter $\Lambda$ satisfies the constraint
\beq\label{lambda_max}
\Lambda\lesssim 1.6\,d_{\eta}^{-2/13}\left(\frac{m_{3/2}}{m_{\eta}}\right)^{3/13} M_P = 8\times 10^{-3}\, \tilde{d}_{\eta}^{-2/13}\left(\frac{m_{3/2}}{10^{-15}M_P}\frac{10^{-5}M_P}{m_{\eta}}\right)^{3/13}M_P.
\eeq

If the limit in (\ref{lambda_max}) is satisfied, the universe is never dominated by $z$ oscillations,
and $\rho_z < \rho_r$ at the time of decay. In this case there will be no net entropy production.
Therefore, for $\Lambda\lesssim 10^{-2}M_P$, all the cosmological problems associated with the evolution of the hidden sector are resolved. In particular, no significant amounts of entropy are generated, and any dilution effects of the products of baryogenesis and nucleosynthesis may be neglected.

For completeness, we return to the case that $\Lambda$ is small enough so that
$z$ decay occurs before inflaton decay.
The ratio $R_{dz}/R_{d\eta}$ is smaller than one for
\beq\label{lambda_min}
\Lambda\lesssim d_{\eta}^{-2/5}\left(\frac{m_{3/2}}{m_{\eta}}\right)^{3/5}M_P=  10^{-6}\, \tilde{d}_{\eta}^{-2/5}\left(\frac{m_{3/2}}{10^{-15}M_P}\frac{10^{-5}M_P}{m_{\eta}}\right)^{3/5}M_P.
\eeq
Thus, for smaller $\Lambda$, the decay of the Polonyi field occurs before reheating. In this scenario, the decay occurs when $\Gamma_z=\frac{3}{2}H$, with the Hubble parameter given by eq. (\ref{H1}). The scale factor is given by
\beq
\frac{R_{dz}}{R_{\eta}}=\frac{\pi^{2/3}}{3}\Lambda^{10/3}m_{\eta}^{2/3}m_{3/2}^{-2}M_P^{4/3}.
\eeq
The universe is dominated by the oscillations of the inflaton field, since $\rho_{\eta}/\rho_z = 16(\Lambda/M_P)^{-4}\gg 1$ at $Z$ decay. The entropy release due to the modulus decay is clearly negligible
in this case.


\section{Dark matter production and the gravitino problems}

Having resolved the problem of entropy production, we turn to another of the serious
issues facing moduli in cosmology, namely the overproduction of non-thermal relics.
As we have shown, the Polonyi modulus decays predominantly into a pair of gravitinos.
This implies that the gravitino density produced by $Z$ decay is $n_{3/2}=2n_z$,
where $n_z$, $n_{3/2}$ denote the number density of $Z$ and the gravitino respectively.
In addition, inflation may be a source of gravitinos through direct decay or through
thermal processes during reheating. Gravitinos in turn will decay into an odd number of lightest supersymmetric particles (LSP), provided $R$-parity is a good symmetry.
If the decay of $Z$ into gravitinos is too efficient ($\Lambda$ much smaller than one),
or if the thermal reheat temperature after inflation is too high,  gravitinos may be too copiously produced, and the resulting LSP abundance will be large enough to over-close the universe. We will assume that the LSP corresponds to a neutralino. In this case, direct production of LSPs by the decay of $Z$ is negligible.

The measured cold dark matter density \cite{planck},
assumed to be neutralinos, leads to a direct bound on the
abundance of gravitinos. The closure fraction in neutralinos produced
by gravitino decay, assuming $n_\chi = n_{3/2}$,  can be written as
\beq
\Omega_\chi  \simeq \frac{7 m_\chi n_{3/2} n_\gamma}{s \rho_c} \simeq 2.75 \times 10^9 \left( \frac{m_\chi}{100 {\rm GeV}} \right) \frac{n_{3/2}}{s} ,
\eeq
where $s$ is the entropy density $\simeq 7 n_\gamma$ today and $\rho_c$ is the closure density.
Thus for $\Omega_\chi h^2 \lesssim 0.12$, we have an upper limit
\beq
\frac{n_{3/2}}{s} \lesssim 4.4 \times 10^{-12} \left( \frac{100 {\rm GeV}}{m_\chi} \right)
\label{limit}
\eeq
For sufficiently heavy gravitinos ($m_{3/2} \gtrsim 10$ TeV), this bound dominates
over the limit from big bang nucleosynthesis (see e.g. \cite{ceflos}).

Naively, the direct decay of an inflaton to a gravitino and inflatino could easily violate the bound
(\ref{limit}). If one assumed that the density of gravitinos was equal or close to the
number density of inflatons prior to their decay, the gravitino density would
scale as $n_{\eta}/s \sim (m_\eta/M_P)^{1/2}$. However direct decays may be kinematically
forbidden \cite{nos} if $|m_\eta - m_{\tilde \eta}| < m_{3/2}$ where $m_{\tilde \eta}$ is the mass
of the inflatino, or  is kinematically suppressed \cite{nop} if $m_{3/2} \ll m_{\eta}\simeq m_{\tilde \eta}$.
In that case, $n_{3/2}/s \sim (m_\eta/M_P)^{1/2} (m_{3/2}/m_{\eta})$ and would safely satisfy the bound
(\ref{limit}). We will therefore ignore the direct production of gravitinos from inflaton decay.

The thermal production of gravitinos during reheating is potentially more problematic
as it is proportional to the reheat temperature (\ref{tr}). The gravitino-to-entropy ratio from thermal production is calculated to be \cite{bbb}
\beq\label{thyield}
\frac{n_{3/2}}{s}=2.4\times10^{-12}\left(\frac{T_R}{10^{10}\,{\rm GeV}}\right) =
2.6\times10^{-11}d_{\eta}g_{\eta}^{-1/4}\left(\frac{m_{\eta}}{10^{-5}M_P}\right)^{3/2} ,
\eeq
for $m_{1/2}\ll m_{3/2}$. Combining Eqs. (\ref{limit}) and (\ref{thyield})
we have,
\beq
d_{\eta}g_{\eta}^{-1/4}\left(\frac{m_{\eta}}{10^{-5}M_P}\right)^{3/2}  \left( \frac{m_\chi}{100 {\rm GeV}} \right) \lesssim 0.17 ,
\label{thlimit}
\eeq
which can clearly be satisfied.

Finally, we discuss the abundance of gravitinos computed by determining the number density of $Z$ when it decays. As we will see, the limit on $\Lambda$ from the
non-thermal production of neutralinos is stronger than the limit from entropy production derived above.
Therefore, in this section, we will assume that the bound (\ref{lambda_max}) is satisfied and
$\Lambda$ is sufficiently small so that $Z$ never dominates the energy density.

\begin{figure}[!h]
\centering
	\scalebox{1.0}{\includegraphics{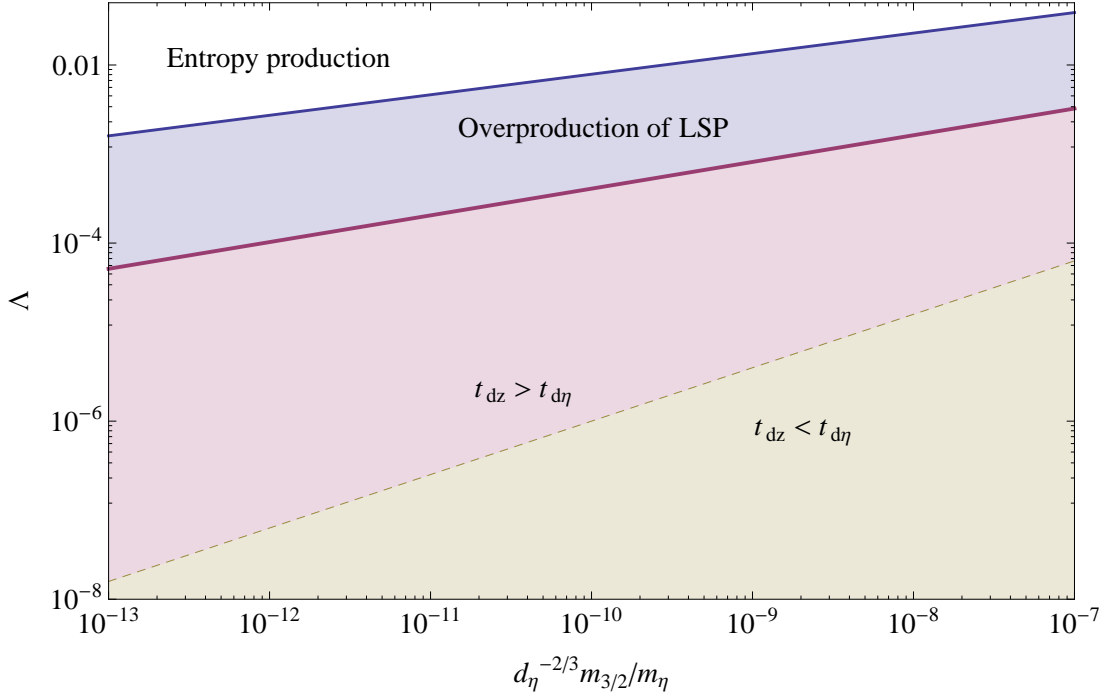}}
	\caption{Allowed range for $\Lambda$  as a function of $d_{\eta}^{-2/3}\,m_{3/2}/m_{\eta}\simeq 1.3g^{-1/6}\,m_{3/2}/(T_{R}^2 M_P)^{1/3}$. The upper limit corresponds to (\ref{lambda_max}), while the boundary at which $Z$ decays at reheating is given by (\ref{lambda_min}).  The purple curve corresponds to eq. (\ref{omega}),  which can be rewritten as $\Lambda=0.075\,(g/g_{\rm SM})^{1/20}(\Omega_{\chi}h^2/0.1199) (m_{\chi}/100\,{\rm GeV})^{-1/5}(m_{3/2}/10^{-15}M_P)^{-1/10}(\tilde{d}_{\eta}^{-2/3}\,m_{3/2}/m_{\eta})^{3/10}$.}
\label{z1}
\end{figure}

The number density of $z$'s is dependent on whether the decay occurs before or after reheating,
\beq
n_z = \frac{\rho_z}{m_z} \simeq
\begin{cases}
0.033\, d_{\eta}\Lambda^{-5/2}m_{\eta}^{3/2}m_{3/2}^{7/2}M_P^{1/2}, & R_{dz}>R_{d\eta}\\
0.066\Lambda^{-5}m_{3/2}^5M_P^3, & R_{dz}<R_{d\eta}
\end{cases}
\eeq
In both cases, the resulting gravitino number density to entropy ratio is given by
\beq\label{yield}
\frac{n_{3/2}}{s}=0.038\, g_{\eta}^{-1/4}d_{\eta}\Lambda^{5}\left(\frac{m_{\eta}^{3/2}}{m_{3/2}M_P^{11/2}}\right).
\eeq
The corresponding neutralino yield is $n_{\chi}/s\simeq n_{3/2}/s$.
Thus, the neutralino density parameter $\Omega_{\chi}=m_{\chi}n_{\chi}/\rho_c$ is evaluated to be
\beq\label{omega}
\Omega_{\chi}h^2 \simeq 0.12\,g_{\eta}^{-1/4}d_{\eta}\left(\frac{\Lambda}{3.2\times 10^{-4}M_P}\right)^{5}\left(\frac{m_{\chi}}{100\,{\rm GeV}}\right)\left(\frac{m_{\eta}}{10^{-5}M_P}\right)^{3/2}\left(\frac{10^{-15}M_P}{m_{3/2}}\right),
\eeq
and the scale $\Lambda$ which provides the necessary strong stabilization for the Polonyi modulus may be tuned to yield a density parameter consistent with the Planck normalization for the dark matter content of the universe \cite{planck}. The value of $\Lambda$ for which the correct relic density is obtained corresponds to the reheating-before-decay scenario (see Figure \ref{z1}), and to the mass $m_{Z}\sim 10^7\ {\rm GeV}$.

The validity of the expression (\ref{omega}) for the density parameter depends on the assumption that no significant amount of entropy is released at the decay of the gravitino. Since gravitinos are weakly interacting and non-thermally produced from the decay of the Polonyi field, they are effectively thermally decoupled until their decay. Gravitinos are relativistic at the time of production, since $m_z\gg m_{3/2}$, but they are slowed down by redshift, with momenta $p\propto R^{-1}$ \cite{redshift}. The dominant decays of the gravitino correspond to decays into a standard model particle and its supersymmetric partner, with rate \cite{moroith}
\beq
\Gamma_{3/2}(\psi_{3/2}\rightarrow {\rm MSSM})\simeq \frac{193}{384\pi} \frac{m_{3/2}^{3}}{M_P^2}.
\eeq
With the scale factor at gravitino decay given by $(R_{3/2}/R_{dz})^2=\Gamma_z/\Gamma_{3/2}\simeq 10.3\Lambda^{-5}\gg 1$, the gravitino will be non-relativistic at the time of decay. Approximating the energy density as $\rho_{3/2}=\rho_z(m_{3/2}/2m_{z}) (R_{dz}/R)^3$, with $\rho_z$ the energy of the Polonyi field at its decay, a straightforward calculation shows that when the gravitino decays, the universe is dominated by the relativistic products of the inflaton, $\rho_r/\rho_{3/2}>1$, if the mass scale $\Lambda$ satisfies $\Lambda\lesssim 2.1\times10^{-3}M_P$. Therefore, the gravitino never dominates the universe, and no significant amount of entropy is released at its decay.

Since the thermal production of gravitinos is independent of $\Lambda$, we
can compare the thermal abundance with that produced by Polonyi decays.
The ratio of the gravitino yield produced by modulus decay (\ref{yield}) to the thermally produced yield (\ref{thyield}) is
\beq
\frac{(n_{3/2}/s)_{Z\ {\rm decay}}}{(n_{3/2}/s)_{\rm thermal}} \simeq 4.58\times10^{16}\left(\frac{\Lambda}{M_P}\right)^5\left(\frac{10^{-15}M_P}{m_{3/2}}\right).
\eeq
Using this comparison and assuming that we satisfy the bound (\ref{thlimit}), we can obtain a bound on
$\Lambda$ which insures that gravitinos gravitinos prodiuced by Polonyi decay is subdominant.
This is the case if
\beq
\Lambda\lesssim 4.7 \times10^{-4}M_P \left(\frac{m_{3/2}}{10^{-15} M_P}\right)^{1/5} .
\eeq

Finally, if we include the effects of annihilations, the neutralino abundance produced by gravitino decay is determined from the Boltzmann equation
\beq
\frac{dn_{\chi}}{dt} + 3Hn_{\chi} = -\langle \sigma_{\rm ann} v_{\rm rel}\rangle n_{\chi}^2.
\eeq
where $\langle \sigma_{\rm ann} v_{\rm rel}\rangle$ denotes the thermal-averaged annihilation cross section.
If the universe is dominated by the energy density of radiation, $\rho_{r}>\rho_{\rm LSP}$, and since the entropy release from the gravitino decay is negligible, the relic abundance is found to be  \cite{jeong,fujii,wino}
\beq
\left(\frac{n_{\chi}}{s}\right)^{-1}\simeq \left(\frac{n_{\chi}}{s}\right)_{3/2}^{-1} + \left(\frac{H}{s\langle \sigma_{\rm ann}v_{\rm rel}\rangle}\right)^{-1}_{3/2},
\eeq
where the subindex indicates evaluation at gravitino decay. Therefore, the previous result (\ref{omega}) is only altered if the annihilation term is smaller than the gravitino yield (\ref{yield}). For typical annihilation rates for neutralino LSP, $\langle\sigma_{\rm ann}v_{\rm rel}\rangle \sim 10^{-7}-10^{-8}\,{\rm GeV}^{-2}$ \cite{myy, wino, KSWY}, the ratio
\[
\left(\frac{H/s\langle \sigma_{\rm ann}v_{\rm rel}\rangle}{n_{\chi}/s}\right)_{3/2}\sim \frac{10^4}{d_{\eta}}\left(\frac{\Lambda}{3\times10^{-4}M_P}\right)^{-5} \left(\frac{m_{\eta}}{10^{-5}M_P}\right)^{-3/2}\left(\frac{m_{3/2}}{10^{-15}M_P}\right)^{-1/2} \left(\frac{10^{-7}\,{\rm GeV}^{-2}}{\langle\sigma_{\rm ann}v_{\rm rel}\rangle}\right)
\]
is much larger than one, indicating that pair annihilation of neutralinos is not effective, and all the produced LSP's during gravitino decay survive. 

\section{Summary and conclusion}

We have considered the cosmological consequences of a  strongly stabilized, supersymmetry breaking hidden sector. The degree of stabilization is characterized by a mass scale, $\Lambda$, defined in the
K\"ahler potential.
We have shown that solutions to the cosmological problems inherent to light moduli in supergravity
are possible for sufficiently small $\Lambda$. In this approach, the Polonyi sector is not responsible for providing the reheating temperature necessary for nucleosynthesis, since its energy density and the entropy released by its decay are subdominant with respect to that of the inflaton field. This restriction could easily be relaxed in some scenarios of the Affleck-Dine mechanism of baryogenesis where the late entropy release from the modulus decay is necessary to dilute a large baryon asymmetry. Nevertheless, a large baryon asymmetry is not a generic feature of the Affleck-Dine mechanism, and a negligible entropy release from modulus decay is in some cases necessary to obtain an asymmetry consistent with observations. This is true in particular when the flat direction responsible for the asymmetry is lifted by non-renormalizable quartic operators in the superpotential \cite{DRT, gmo,ADinf}.

Our results are neatly summarized in Figure \ref{z1} which shows the various physical
regimes discussed above for $\Lambda$ as a function of the dimensionless combination $d_\eta^{-2/3} m_{3/2}/m_{\eta}$. At large $\Lambda$, there is an excessive amount of entropy produced
as in the classic Polonyi scenario. At somewhat lower $\Lambda$, although the Polonyi
field never comes to dominate the energy density of the universe, its decay leads to the over-production
of the LSP. The figure also demarcates the values of $\Lambda$ such that the Polonyi field
decays before or after the inflaton. The figure does not show, however,
the additional constraint (\ref{thlimit}) derived from thermally produced gravitino decay
as this constraint is independent of the dynamics of the Polonyi sector.

It must be emphasized that the introduction of the single stabilizing parameter $\Lambda$ not only accounts for the solution of the entropy problems related to the Polonyi field, but it may also preclude the later onset of a gravitino and neutralino problem from moduli decay. Unless the LSP is copiously produced during inflaton decay or by scatterings in the primordial plasma, the suppression of all decay channels of the hidden sector relative to the gravitino channel imply that the bulk of the relic LSP density is generated from the decay of the gravitinos produced by the modulus decay. In this sense, the decay of the strongly stabilized Polonyi field can account for the present dark matter abundance. The constraint on $\Lambda$ coming from the observed abundance, $\Omega_{\chi}h^2=0.1199$, lies well within the bound imposed by the resolution of the cosmological problems for the Polonyi field.

\section*{Acknowledgments}

We would like to thank A. Linde and T. Yanagida for helpful discussions.
This work was supported in part
by DOE grant DE--FG02--94ER--40823 at the University of Minnesota.


\end{document}